\documentstyle{mn}
\begin{document}

\def\lsun{{L_{\odot}}}
\def\msun{{\rm M_{\odot}}}
\def\rsun{{R_{\odot}}}
\def\be{\begin{equation}}
\def\ee{\end{equation}}
\def\m2i{M_{2,\rm i}}
\newcommand{\lta}{\la}
\newcommand{\gta}{\ga}
\input{epsf.sty}
\def\plotone#1{\centering \leavevmode
\epsfxsize=\columnwidth \epsfbox{#1}}
\def\plottwo#1#2{\centering \leavevmode
\epsfxsize=.45\columnwidth \epsfbox{#1} \hfil
\epsfxsize=.45\columnwidth \epsfbox{#2}}
\def\plotfiddle#1#2#3#4#5#6#7{\centering \leavevmode
\vbox to#2{\rule{0pt}{#2}}
\includegraphics{#1}}

\title{The CV period gap: still there}
\author[U. Kolb, A. R. King, \& H. Ritter]{U. Kolb$^{1,2}$,
A. R. King$^1$ \& H. Ritter$^{2}$\\ 
$^1$ Astronomy Group, University of Leicester,
Leicester, LE1~7RH\\ $^2$ Max--Planck--Institut f\"ur Astrophysik, 
Karl--Schwarzschild--Str. 1, D--85740 Garching, Germany}

\date{May 1998 \\ MNRAS, in press}

\maketitle

\begin{abstract}
We consider a recently--proposed alternative explanation of the CV
period gap in terms
of a revised mass--radius relation for the lower main sequence. We
show that no such thermal--equilibrium relation is likely to
produce a true gap. 
Using population synthesis techniques we calculate a model population
that obeys the claimed equilibrium mass--radius relation. A
theoretical period histogram obtained from this population shows
two prominent period spikes rather than a gap. 
We consider also recent
arguments suggesting that the period gap itself may not be
real. We argue that, far from demonstrating a weakness of the
interrupted--braking picture, the fact that most CV subtypes prefer
one side of the gap or the other is actually an expected consequence
of it. 
\end{abstract}

\section{Introduction}
\label{sec:intro}
The dearth of cataclysmic variables (CVs) between $\sim 2$ and $\sim
3$~hr is well
known as the CV period gap. The conventional explanation of
this (see e.g. King, 1988; Kolb, 1996; Ritter, 1996 for reviews)
invokes the idea that mass transfer on a timescale shorter than the
secondary's thermal time will cause this star to be larger than its
thermal--equilibrium radius. If there is a sudden drop in the orbital
braking once the secondary becomes fully convective, it is plausible
that a gap of the observed width results. The lack of a cogent
predictive theory of the braking mechanism for CVs above the gap means
that no explanation of this kind can be totally compelling. However,
the interrupted--braking picture does naturally explain several
features such as the sharpness of the gap sides, and (to some extent)
the distributions of various CV subtypes across it. It has shown
remarkable resilience in withstanding various alternative suggestions
(e.g. Livio \& Pringle, 1994 -- see King \& Kolb, 1995).

In a recent paper, Clemens et al.\ (1998; hereafter C98) 
suggest another alternative theory for the formation
of the gap. This is based on a proposed revision of the mass--radius
relation for the lower main sequence, as derived 
from colour--magnitude diagrams of nearby stars (Reid \& Gizis,
1997). C98 derive radii and masses using bolometric corrections and
the temperature scale from Leggett et al.\ (1996), and an empirical
mass--magnitude relation (Henry \& McCarthy, 1993).
C98 then assume that CV secondaries do not significantly deviate from
this equilibrium mass--radius relation. Thus the effective
mass--radius index 
\be
\zeta = \frac{{\rm d} \ln R_2}{{\rm d} \ln M_2} 
\label{zeta}
\ee
($R_2$, $M_2$ are the donor mass and radius)
along the evolution above and through the period gap
region must always equal the local equilibrium index of the main
sequence. 
In C98's interpretation, $\zeta$ takes values close to
1/3 in two different sections of the main sequence. Any range of 
secondary masses $M_2$ with $\zeta = 1/3$ maps on to a single binary
period (see eqs.~(\ref{P}), (\ref{Pdot}) below), in this case to about
3.4~hr and 2.0~hr.  
C98 show that this gives a rather slow evolution
in period at these two points, and claim that this reproduces the CV
period gap.

We shall not comment on C98's interpretation of these data, except to
say that it is certainly not the only possible one. Our main point is
to take issue with C98's conclusion that their revision allows
an alternative explanation for the period gap. We shall show that, far
from producing a period gap, this revision in fact produces a pair of
period `spikes' at 3.4~hr and 2.0~hr. The CV discovery probability
between these two spikes is no lower than outside them. We derive the
condition for a mass--radius relation to
produce a gap, and point out that this is very unlikely to be
satisfied by any thermal--equilibrium mass--radius relation.
Finally we briefly consider the argument by Verbunt (1997) questioning
the reality of the period gap.

\section{Secular evolution with a fixed mass--radius relation}

Roche geometry allows one to express the orbital period change 
in terms of $\zeta$. The secondary's Roche lobe radius can be
approximated as
\be
R_{\rm L} = C\biggl({M_2\over M}\biggr)^{1/3}a
\label{RL}
\ee
(Paczy\'nski, 1971), where $M$ is the total binary mass, $C \simeq
0.462$ and $a$ is the binary separation. Combining this with Kepler's
3rd law shows that the binary period is given by
\be
P \propto \biggl({R_{\rm L}^3 \over M_2}\biggr)^{1/2}. 
\label{P}
\ee
From (\ref{zeta}) we have 
\be
{\dot R_2\over R_2} = \zeta{\dot M_2\over M_2},
\label{R2dot}
\ee
hence with (\ref{P}) for the rate of change of period 
\be
{\dot P\over P} = {3\zeta -1 \over 2}{\dot M_2\over M_2}.
\label{Pdot}
\ee
We thus see that $\dot P$ vanishes where $\zeta = 1/3$.

The binary evolves in response to the loss of orbital angular
momentum $J$. This may occur through gravitational radiation or other
processes, e.g. magnetic braking. 
With primary mass $M_1$ we can write
\be 
J = M_1M_2\biggl({Ga\over M}\biggr)^{1/2}.
\label{J}
\ee
In the following we shall assume that all the mass lost by the
secondary is accreted by the primary, so that the total binary mass
remains fixed, i.e. $\dot M_1 = -\dot M_2 > 0, \dot M = 0$. Then
logarithmically differentiating (\ref{RL}, \ref{J}) gives
\be
{\dot R_{\rm L}\over R_{\rm L}} = {2\dot J\over J} - 
{2\dot M_2\over M_2}\biggl({5\over 6} - {M_2\over M_1}\biggr).
\label{RLdot}
\ee
For stationary mass transfer (i.e. $\ddot M_2 \simeq 0$) the Roche
lobe and stellar radius must move in step, i.e. $\dot R_{\rm L}/
R_{\rm L} = \dot R_2/R_2$. With (\ref{R2dot}) and (\ref{RLdot}) this
gives the mass transfer rate as
\be
{\dot M_2\over M_2} = {\dot J/J\over 5/6 + \zeta/2 - M_2/M_1}.
\label{Mdot}
\ee
The denominator of this equation is automatically positive if we
require mass transfer stability, i.e. that in the absence of angular
momentum losses the Roche lobe would expand away from the star 
($\dot R_{\rm L}/R_{\rm L} > \dot R_2/R_2$) in
response to mass transfer. This follows from setting $\dot J = 0$ in
(\ref{RLdot}) and comparing with (\ref{R2dot}), and in general requires
that the mass ratio $M_2/M_1$ should be $\lta 1$.

Thus, specifying $\dot J$ in some way, the binary evolution is
fixed. In particular we can express the mass transfer rate $-\dot M_2$ and
period derivative $\dot P$ as functions of the secondary mass
$M_2$. The probability $p(\log P)$ of discovering the system
in a given range of $\log P$ will vary as 
\be
p(\log P) \propto {|-\dot M_2|^{\alpha}\over |\dot P/P|}
\label{p},
\ee
where $\alpha$ is some positive power. For a
bolometrically flux--limited sample 
drawn from an isotropically distributed population 
we have $\alpha = 3/2$. Power--law
fits to the bolometric correction show that $\alpha \simeq 1$ is appropriate
for a visual magnitude--limited ($m_{\rm vis}$--limited) sample (see
below), and indeed this value gives a better fit to the observed
distribution in the conventional picture (Kolb 1996).

From (\ref{Pdot}, \ref{Mdot}) we get
\be
p(\log P) \propto M_2^{\alpha}\biggl|{\dot J\over
J}\biggr|^{\alpha - 1}(3\zeta - 1)^{-1}\biggl({5\over 6} + {\zeta \over
2} - {M_2\over M_1}\biggr)^{1 - \alpha}.
\label{p(M)}
\ee
For the particular choices $\alpha = 3/2$ and $1$ we have
\be
p(\log P) \propto M_2^{3/2}\biggl|{\dot J\over
J}\biggr|^{1/2}(3\zeta - 1)^{-1}\biggl({5\over 6} + {\zeta \over
2} - {M_2\over M_1}\biggr)^{-1/2}
\label{3/2}
\ee
and
\be
p(\log P) \propto \frac{M_2}{3\zeta - 1}.
\label{1}
\ee

\section{The theoretical period histogram}

\begin{figure}
\centerline{\plotone{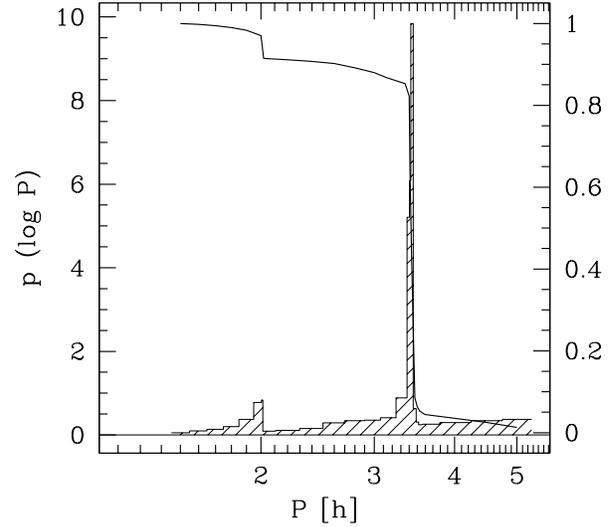}}
\caption{
Period histogram $p(\log P)$ (hatched, scale on the left)
predicted by the proposed mass--radius
relation of C98, with the cumulative
distribution superimposed (solid curve, scale on the right), for CVs
with $P < 5$~hr.  
The figure assumes $\alpha = 1$ in eqn.~(\protect\ref{p}), as
appropriate for a (visual) magnitude--limited sample, and was obtained
with $-\dot M_2$ and $\dot P/P$ taken from C98.
\label{pdista}
}
\end{figure}

The last three equations show us the effect of any choice $\zeta(M_2)$
on the discovery probability, which is the quantity predicting the CV period
histogram. Since in general $M_2$ and $|\dot J/ J|$ are smooth
functions through the evolution, the character of the
predicted histogram is controlled by $\zeta$. 
The choice of $\zeta(M_2)$ given by C98 has $\zeta \simeq 1/3$ for
secondary mass ranges which translate to periods 3.4 and 2.0~hr. 
But from (\ref{p(M)}) we see that this actually makes $p \rightarrow
\infty$ at these two points 
rather than reducing $p$ towards zero between them, as required
for a period gap. 
Fig.~\ref{pdista} shows $p(\log P)$ for the case $\alpha = 1$, evaluated directly
from (\ref{p}) with $-\dot M_2$ and $\dot P/P$ from C98.
As can be seen, this does not give a period gap at
all, but two spikes. The small period derivative at $P=3.4$~hr
raises the discovery probability dramatically there, as the systems
spend a long time near this period without being particularly
faint there. A similar but weaker effect is seen at $P=2.0$~hr. 
Note
that the discovery probability between the two spikes is {\it not}
lower than that outside them. The case $\alpha = 3/2$ (eq. \ref{3/2})
gives similar results, but with even larger spikes. 
Neither of these
distributions looks at all like the observed one (see Fig.~\ref{obs}
below), which has roughly equal numbers above and below the gap, with
approximately uniform distributions in $\log P$ in each case.  
The disagreement with observation is even worse for $\alpha=3/2$.

The discovery probability (\ref{p}) refers to a particular CV system 
at period $P$ with current mass transfer rate $\dot M$ and
corresponding period derivative $\dot P$. The observed period
histogram however represents a sample of systems
selected from the present--day Galactic CV population according to
certain selection effects. Thus the sample comprises systems  
with different ages at different evolutionary stages. In the next
section we make use of the population synthesis technique developed by
Kolb (1993) and de~Kool (1992) to synthesize a theoretical observable
period distribution in the framework of a given evolutionary model for
CVs. The key element in this procedure is to enforce the mass--radius
relation as claimed by C98 on the lower main sequence. As we will see
the resulting period distribution does indeed show similar spikes as
the discovery probability in Fig.~\ref{pdista}, and no gap.

\subsection{A population model}

CVs are believed to form from a subset of zero--age binaries with a 
low--mass companion which undergo a common envelope (CE) phase (see
e.g.\ de~Kool 1992, Politano 1996).  
The CE evolution tightens the orbit and exposes the white dwarf core
of the giant primary. Orbital angular momentum losses $\dot J$ further
shrink the orbit of the post--CE binary until the secondary fills its
Roche lobe. The subsequent semi--detached evolution is governed by
$\dot J$ and the secondary's response $\zeta$ to mass loss, as
outlined in Sect.~2.   

The method for obtaining a full theoretical period histogram has been
described extensively by Kolb (1993).  
Chiefly, there are three main steps involved: 
first, to calculate the time--dependent formation rate of
nascent CVs, e.g.\ as a function of initial white dwarf and donor mass,
for given formation rate and distributions of primary mass, mass 
ratio and orbital separation of zero--age binaries.
Second, to evolve this nascent population into its
present--day configuration. The resulting multi-dimensional
distribution function defines the intrinsic present--day Galactic CV
population.  
Third, selection effects have to be defined which determine
how the observable sample is drawn from the intrinsic population.

In C98's attempt to explain the CV period gap the nature of the
angular momentum loss $\dot J$ is irrelevant as long as it is
continuous and smooth along the evolution. Obviously, for a 
population synthesis we have to specify $\dot J$. We adopt the
simplest choice for a continuous $\dot J$ and assume that it is given
by the emission of gravitational waves only. 
For this case the time--dependent CV fomation rate has already been
calculated by Kolb \& de~Kool (1993), using a Monte Carlo technique
(see also de~Kool 1992). In their simulations the primary and
secondary of zero--age binaries form independently from a Miller \&
Scalo type IMF (Miller \& Scalo 1979), the distribution of initial
orbital separations $a$ is flat in $\log a$, and the common envelope
efficiency is 1.     

\begin{figure}
\centerline{\plotone{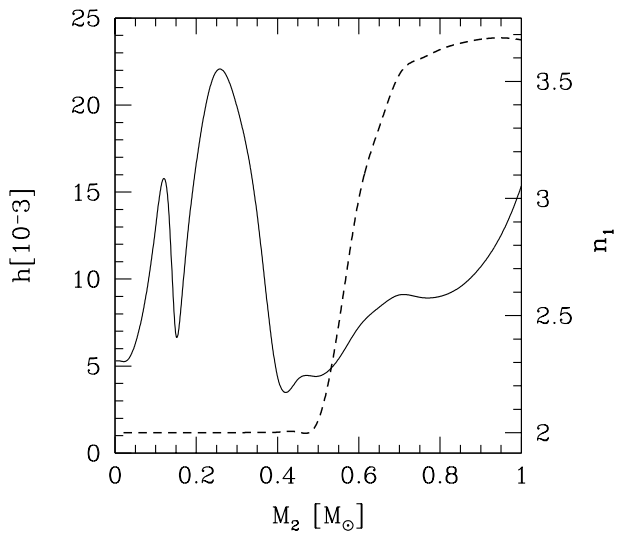}}
\caption{
Bipolytrope calibration functions (solid: entropy jump $h$; dashed:
core index $n_1$), used to mimic the mass--radius
relation claimed by C98. 
\label{cal}
}
\end{figure}

To calculate the secular evolution of CVs we
use a simplified stellar structure code that describes 
the donor's radiative core (if present) as a polytrope with index
$n_1$, the donor's convective envelope as a polytrope with index
$1.5$. 
The bipolytrope model has two free parameters, the core index $n_1$ and a
discontinuity $h$ of the specific entropy at the stellar surface (see
Kolb \& Ritter 1992). We calibrate these parameters as a 
function of donor mass (Fig.~\ref{cal}) in such a way that the
corresponding (bi)polytrope models in thermal equilibrium reproduce
the C98 mass--radius relation to better than $1\%$ in the range
$0.12<M_2/\msun<0.65$. For larger masses we used the same calibration
as in Kolb \& Ritter (1992), for smaller masses a calibration which
reproduces the CV minimum period at $\simeq 80$~mins. 
To mimic the effect of nova outbursts we invoke an isotropic wind from
the white dwarf, which removes the transferred material with the primary's
specific orbital angular momentum from the system. 
We note that this stellar evolution code self--consistently takes into
account the donor's deviation from thermal equilibrium as a result of
mass transfer, i.e.\ does not force $\zeta$ to be {\it equal} to the
equilibrium index along an evolutionary sequence. As $\dot J$ is
rather weak the actual deviations are of course small. 

\subsection{Results}

\begin{figure}
\centerline{\plotone{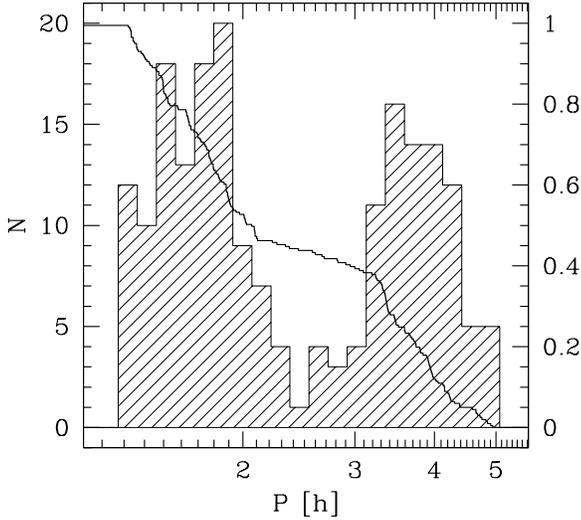}}
\caption{
Period histogram (hatched; scale on the left) and cumulative
distribution (solid curve; scale on the right) for observed CVs with $P <
5$~hr. Data taken from Ritter \& Kolb (1998).
\label{obs}
}
\end{figure}

\begin{figure}
\centerline{\plotone{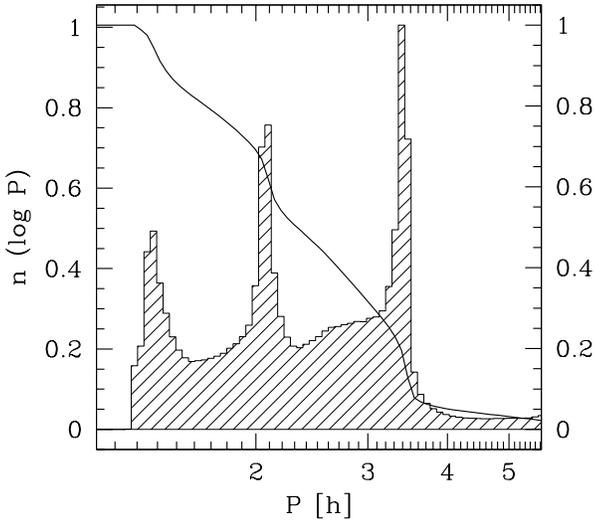}}
\caption{
As Fig.~\protect{\ref{obs}}, but for a visual magnitude--limited sample
drawn from a model population obeying the equilibrium mass--radius
relation claimed by C98.
\label{popsyn}
}
\end{figure}

In Fig.~\ref{popsyn} we plot the period histogram $n(\log P)$ of a
$m_{\rm vis}$--limited sample taken from the resulting intrisic 
model population. To do so we neglected the drop of the CV space
density perpendicular to the Galactic plane and assumed that the visual
light comes from an optically thick, steady-state accretion disc. In
this case the integration of the intrinsic space density distribution
over the observable volume can be approximated by a multiplication
with the differential selection factor $s \propto 
L_{\rm vis}^{3/2} \propto M_1^{3/2} \dot M^{9/8}$ (cf.\ Kolb 1993,
Kolb 1996).   

As expected, the period distribution in Fig.~\ref{popsyn} shows two
pronounced peaks at $P\simeq2$~hr and $3.4$~hr. The population density
$n$ between these spikes is not lower than outside of them, i.e.\
there is no gap. The drop of $n$ for $P\ga3.5$~hr is from a drop in
the formation rate 
at larger donor mass, and is also seen in the period distribution
obtained by Kolb \& de~Kool (1993) with the more conventional underlying
main--sequence mass--radius exponent $\simeq 0.8 \simeq$~constant.   
The third spike at the minimum period $\simeq 1.3$~hr occurs for the
same reason as the other two spikes, i.e.\ because $\zeta=+1/3$. It 
indicates the transition from the lower main sequence to the brown
dwarf regime where $\zeta \rightarrow -1/3$.     
It is well--known that the observed period histogram (Fig~\ref{obs})
has no such spike at the lower period cut--off. This suggests that
additional selection effects prevent us from detecting these extremely
low--mass donor CVs.

Changing the exponents of $M_1$ and $\dot M$ in $s$, as would be
appropriate for samples that are not purely $m_{\rm vis}$ limited,
does not alter the qualitatitive features of the resulting period
histogram. Hence the detailed population synthesis fully confirms the
expectations from the functional of the discovery probability 
(\ref{p(M)}): the mass--radius relation claimed by C98 does generate a
gap in the period distribution of CVs.

From (\ref{p(M)}) it is easy to see that a real period gap -- a region
where $p(\log P)$ is low or zero -- requires a very large effective
$\zeta$. This in 
turn means that we require the secondary's radius to change
significantly for a very small change of its mass. This is indeed
precisely what happens in the conventional explanation of the gap:
$R_2$ shrinks from a non--equilibrium value towards the main--sequence
radius appropriate to the secondary's current mass $M_2$, with {\it
no} change in the latter. Since $P \propto (R_{\rm L}^3/M_2)^{1/2}$ this
radius change implies that the period must change if the star is to
fill its Roche lobe, and thus transfer mass. It is hard to see
how any revision of the {\it equilibrium} mass--radius relation, as
proposed by C98, could have the required property of producing a very
large $\zeta$ for systems in the period gap.

The fact that features of the function $\zeta(M_2)$ describing the
lower main sequence should reappear as a moderate anomaly in the
observed period histogram 
was already noted by Kolb (1993). The theoretical main sequence has a
well--established hump in $\zeta(M_2)$ in the
mass range $0.4 \la M_2/M_{\sun} \la 0.7$, caused by the dissociation
of $H_2$ molecules in the outermost stellar layers. This should
produce a 
characteristic plateau and shoulder in $p(\log P)$ at $P \simeq
5$~hr. This period range is not shown in Fig.~\ref{obs}; the data are still
too sparse at these periods to confirm the reality of this feature.

\section{Is the period gap real?}

We have so far assumed that the period gap is genuine. 
Verbunt (1997) argues that
that this may not be the case, essentially because not all CV
subtypes show a significant gap. 
In particular, almost all (non-magnetic) novalike systems have periods
above the gap, while dwarf novae populate both sides of it.
Verbunt concedes that the period gap is statistically significant for
the dwarf novae, but he argues that the gap might not be significant
if one broke up the dwarf novae into subtypes such as the SU UMa  
and U Gem systems. 
Almost all of the SU UMa systems are below the gap,
the U Gem systems above (for the period distribution of all CV
subtypes see the catalogue of Ritter \& Kolb, 1998).  

A critical re--evaluation of Verbunt's arguments has already been
given by Hellier \&  Naylor (1998).  They convincingly dismissed his
classification of VY Scl stars with dwarf novae and pointed
out the importance of treating magnetic systems separately for a 
comparison of dwarf novae and novalike systems. 
Here we add to this by emphasizing that Verbunt's argument actually
works 
positively in favour of the conventional theoretical framework of the
explanation of the period gap. 
For the most basic feature of this explanation is that average mass
transfer rates above the gap should be significantly higher than those
below. Thus below the gap we expect that these rates will be too low
to ensure the stability of the accretion discs in non--magnetic
systems, since the disc edge will be too cool to keep hydrogen
ionized (cf Kolb, 1996). 
The prevalence of dwarf novae below the period gap is thus a
natural consequence of the interrupted--braking picture and disc
instability models (see e.g.\ Meyer--Hofmeister \& Ritter, 1993,
for a more detailed discussion).

The coexistence of dwarf novae and novalike (steady) systems at
similar periods above the gap shows that there is evidently a range of
mass transfer rates at a given period. 
The origin of this range is unclear, but it may result from long--term
cycles of the mass transfer rate about the secular mean, possibly
caused by the weak irradiation of the secondary star by the primary
(cf King et al., 1995, 1996). The mean mass transfer rates below the
gap are so far below the critical rate for disc instability that such
fluctuations 
evidently have amplitudes too small to produce many steady systems
below the gap. However above the gap the mean transfer rate and the
critical rate are quite close (cf Kolb, 1996), so the
fluctuations can carry systems across the latter. Thus both
outbursting (dwarf nova) and persistent (novalike) CVs are found above
the gap.

More generally, since the mass transfer rate $-\dot M_2$ and the
mass ratio $M_2/M_1$ are the two most
important parameters characterizing the immediate appearance of a given
CV system, it would be surprising if the explanations for various
types of CV behaviour did {\it not} involve them. Since both quantities
are systematically smaller below the gap than above, one would
thus expect a preference for one side of the gap or the other for
various subtypes, as is actually seen in dwarf novae.

In any case, Verbunt's argument cannot be regarded as compelling,
since it signally fails to provide a reason other than 
unspecified ``selection effects''
for the very strongly significant presence of a period gap when one
takes the full 
set of systems in which a white dwarf accretes from a low--mass
companion, i.e. the full CV sample: compare Fig.~\ref{obs}.

\section{Conclusions}

We conclude that the conventional explanation of the CV period gap in
terms of interrupted orbital braking is still the most plausible. In
conjunction with other ideas, such as the disc instability picture, it
offers a simple framework for understanding the period distributions of the
various CV subtypes. This cannot be said of any proposed alternative.

\paragraph*{Acknowledgments} 
ARK thanks the U.K.\ Particle Physics and Astronomy 
Research Council for a Senior Fellowship. Theoretical astrophysics
research at Leicester is supported by a PPARC Rolling Grant.

\end{document}